\documentclass[journal]{IEEEtran}
\ifCLASSINFOpdf
\else
\fi
\usepackage{array}

\ifCLASSOPTIONcompsoc
  \usepackage[caption=false,font=normalsize,labelfont=sf,textfont=sf]{subfig}
\else
  \usepackage[caption=false,font=footnotesize]{subfig}
\fi
\usepackage{todonotes}
\usepackage{etoolbox}
\usepackage{url}
\usepackage{multirow}
\usepackage{diagbox}

\newtoggle{hidecomments} 

\iftoggle{hidecomments}{ 
	\newcommand{\adrien}[1]{}
	\newcommand{\naz}[1]{}
	\newcommand{\berat}[1]{}
	\newcommand{\serhat}[1]{}
	\newcommand{\comment}[1]{}
	\newcommand{\tom}[1]{}
}{
	\newcommand{\adrien}[1]{\todo[color=blue!40, inline]{\footnotesize{Adrien: #1}}}
	\newcommand{\tom}[1]{\todo[color=red!40, inline]{\footnotesize{Tom: #1}}}
	\newcommand{\naz}[1]{\todo[color=green!40, inline]{\footnotesize{Naz: #1}}}
	\newcommand{\berat}[1]{\todo[color=yellow!40, inline]{\footnotesize{Berat: #1}}}
	\newcommand{\serhat}[1]{\todo[color=gray!40, inline]{\footnotesize{Serhat: #1}}}
	
}
\graphicspath{{Figures/}}

\begin{document}
%
\title{Exploration of Interaction Techniques for Graph-based Modelling in Virtual Reality}
%
%
%

\author{Adrien Coppens $^{(1)}$, Berat Bicer $^{(2)}$, Naz Yilmaz $^{(3)}$ and Serhat Aras $^{(2)}$\\
$^{(1)}$ Software Engineering Lab, University of Mons, Mons, Belgium\\
$^{(2)}$ Department of Computer Engineering, Bilkent University, Ankara, Turkey\\
$^{(3)}$ Cognitive Science Program, Bogazici University, Istanbul, Turkey\\
{\tt\small Corresponding author: adrien.coppens@umons.ac.be}
}

\markboth{ENTERFACE'19, JULY 8TH - AUGUST 2ND, ANKARA, TURKEY}%
{ENTERFACE'19, JULY 8TH - AUGUST 2ND, ANKARA, TURKEY}
%



\maketitle

\begin{abstract}

Editing and manipulating graph-based models within immersive environments is largely unexplored and certain design activities could benefit from using those technologies.
For example, in the case study of architectural modelling, the 3D context of Virtual Reality naturally matches the intended output product, i.e. a 3D architectural geometry. Since both the state of the art and the state of the practice are lacking, we explore the field of VR-based interactive modelling, and provide insights as to how to implement proper interactions in that context, with broadly available devices. We consequently produce several open-source software prototypes for manipulating graph-based models in VR.

\end{abstract}


%
\IEEEpeerreviewmaketitle

\section{Introduction}
\label{sec:intro}


The overall goal of this eNTERFACE'19 project is to explore multi-modal interactions for manipulating graph-based models, i.e. visual models whose basic structure can be represented in the form of graphs, in Virtual Reality (VR).
To approach the problem with different views, our team is composed of people with different backgrounds (computer science, computer engineering, architectural design and cognitive science).

We chose to work on architectural design as a case study, more specifically on parametric modelling. Recent work \cite{coppens2018parametric} has enabled ``mesh streaming" from Grasshopper\footnote{\url{https://www.grasshopper3d.com}}, a popular parametric modelling tool, to Virtual Reality, and identified benefits or visualising a geometry in such a context.

\begin{figure*}[!t]
\centering
\subfloat[Graph]{\includegraphics[width=4.2in]{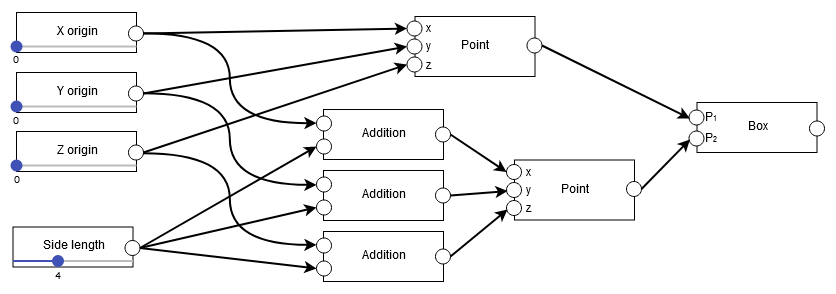}%
\label{fig:simple-cube-graph}}
\hfil
\subfloat[Geometry]{\includegraphics[width=1.1in]{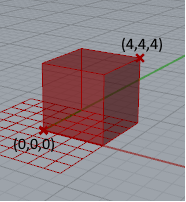}%
\label{fig:simple-cube-viz}}
\caption{DAG (a) for a simple parametric model defining a cube, with the corresponding output geometry (b).}
\label{fig:simple-cube}
\end{figure*}

When designing with Grasshopper, an architect works in a visual programming language that relies on models based on directed acyclic graphs (DAGs), as an underlying representation.
In such a graph, edges contain either standard parameters (e.g. numbers, booleans) or geometries that nodes process in order to output other geometries. The generated result can then flow into the graph and be used as input for other nodes. A complete architectural model can be designed that way, with the intended final geometry being typically output by a sink node. 
Figure \ref{fig:simple-cube} depicts such a DAG for a simple parametric model, whose purpose is to draw a cube defined by 2 corner points, $P_1$ at the origin and $P_2$ at $(4,4,4)$ (since the same ``side length" parameter is used for all 3 additions). The resulting geometry (the output of the sink node i.e. ``Box") is shown on the side of that figure.


While \cite{coppens2018parametric} allows the user to modify parameter values within the VR environment, we here enable full editing of the graph-based model i.e. users can add, remove or move nodes, and add or remove edges from it. The resulting changes to the model can be saved back to the original Grasshopper-specific format, thanks to a framework that was developed prior to the workshop.

Although our work focuses on such architecture-oriented models, (most of) our findings will apply to other contexts where working within a VR environment makes sense and where graph-based modelling is required. The interaction techniques we rely on are not even limited to DAG-based models, and examples of application domains include software engineering (e.g. several types of UML diagrams), transportation network design, robotics (e.g. path planning) and the entertainment industry (e.g. scene graphs for games, animation design). These fields are, at least potentially, dealing with 3D content and could therefore benefit from an immersive design environment, that matches the content's dimensionality (e.g. visualising a 3D animation in VR while designing it from the same immersive environment, seems beneficial).

%
%
%
%



\section{Related work}
\label{sec:related}

\subsection{Context of the case study: architectural modelling}
\label{sec:parametric}

\begin{figure*}[tb]
	\centering
	\includegraphics[width=.76\textwidth]{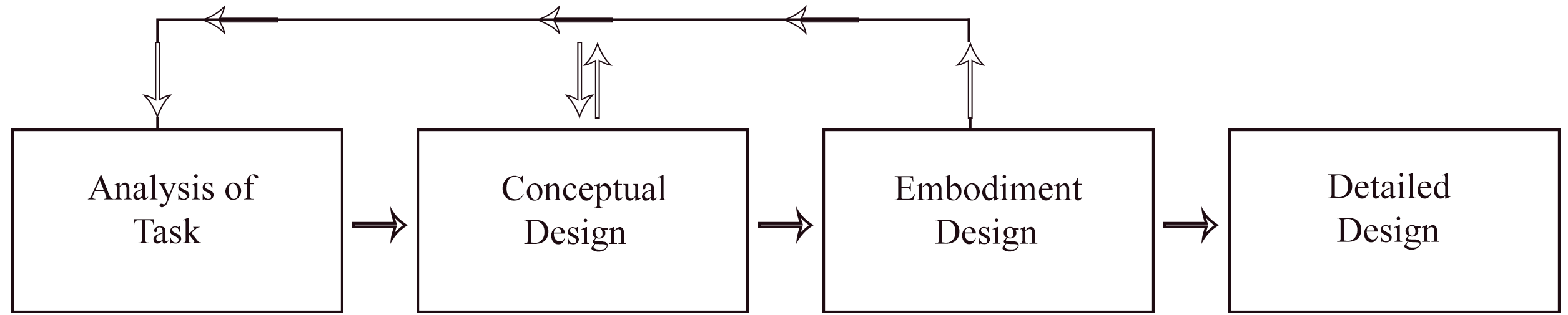}
	\caption{The creative design process as described by Howard et al. \cite{howard2008describing}.}
	\label{fig:design-process}
\end{figure*}

Architectural design tools evolved over the course of the discipline's history. While paper drawings and scale models are still relevant today, they are now accompanied by 2D and 3D modelling software, with limited support for Augmented Reality (AR) and Virtual Reality (VR).
Regardless of the exact tools in use, design activities tend to follow a well-defined process that goes from task analysis to the final product. Howard et al. described such a process \cite{howard2008describing} by combining literature reviews on both engineering design and cognitive psychology. They describe a creative design process  composed of four steps, as depicted in Figure~\ref{fig:design-process}. After analysing the needs for the task in hand, a designer works on creating (often multiple) conceptual designs. Based on a concept, the designer then builds up a structure (i.e. shapes the abstract concept into a concrete design) in the embodiment phase. As indicated by the arrows in that figure, it is quite common to go back and forth between those three steps. Once satisfied with the result, the designer polishes the geometry and produces communication documents for physically building the model.

Most tools that attempts to bring AR/VR technologies in use for architectural modelling focus on the visualisation part and typically provide limited (e.g. texture switching) ”live editing” features (or none at all) that would allow the VR user to modify the geometry whilst immersed into the virtual environment. Examples of such tools include IrisVR Prospect\footnote{\url{https://irisvr.com/prospect/}} and Twinmotion\footnote{\url{https://www.unrealengine.com/en-US/twinmotion}}. There have been several efforts to provide annotation and sketching capabilities in a VR context, some of which were targeted at architectural design activities (e.g. \cite{dorta2016hyve} that relies on a tablet to control a 3D cursor used for sketching), but those are ``only" conceptual design activities.

More recently, the MARUI plugin\footnote{\url{https://www.marui-plugin.com}} has enabled users to modify models directly within a VR environment, therefore tackling the problem of bringing embodiment/detailed design activities into VR. While this clearly appears to be a big step towards the technology's integration for modelling activities, it does not currently offer parametric modelling capabilities.

In order to fill that void, the work we previously mentioned \cite{coppens2018parametric} describes a proof-of-concept application that enables parameter sharing in addition to geometry streaming. This means that an architect can interact with the parametric model's parameter values and see what the resulting effect is on the geometry, all from within the VR environment.
Although this is a good starting point, there is a clear need to expand the tool's capabilities with full model (graph) editing.




\subsection{3D programming languages}
Numerous 3D programming languages for virtual 3D environments have been designed, most of which rely on three-dimensional dataflow diagrams to define programs. Examples from the early 90’s include CUBE/CUBE-II [4][5], a functional language, and Lingua Graphica [6], who translates from/to C++ code.

While the previous examples were indeed designed with virtual environments in mind, they were never adapted to immersive displays (e.g. VR). On the other hand, Steed and Slater implemented an immersive system \cite{steed1996dataflow} that allows users to define object behaviours whilst immersed, once again through dataflow graphs. The system could be used to design animations or interactive applications, that conveniently also took place within the virtual environment.

More recently but with similar goals in mind, Lee et al. developed an Augmented Reality (AR) system \cite{lee2004immersive} to define the behaviour of scene objects for AR applications. Once again, being able to develop and test a target application concurrently was pointed out as a clear benefit.



Since architectural geometries are three-dimensional, we believe designing them from within a VR system would be beneficial to the architect. Even though we therefore are in a 3D context, parametric models (the graphs) are usually two-dimensional, at least that is how they are laid out in the most popular desktop-based applications.
While we also want to explore solutions to capitalise on the third dimension offered by VR environments, we will do so in a limited fashion (e.g. use a node's height to indicate how close it is to the sink)
rather than turning the parametric models into ``unconstrained" 3D graphs.
Indeed, our VR-based system relies on a table metaphor (i.e. the model to be manipulated is placed on a virtual table) to maintain some consistency with desktop tools. We decided to use this metaphor in order to follow the Information Visualisation community's advice on not making use of unmotivated 3D layouts \cite{elmqvist-blog}.

\subsection{Interaction techniques for 3D environments}
\subsubsection{Direct manipulation}
Interacting efficiently within a 3D immersive environment is likely to require input methods that differ from the traditional mouse and keyboard combination. Manipulation can be subdivided into 4 tasks: selection, positioning, rotation and scaling \cite{LaViola3duserinterfaces2017}. While a complete modelling environment needs to afford proper interaction mechanisms for all of these tasks, we will focus here on selection and positioning, the primary needs for our case study.

One way of categorising immersive interaction techniques is based on isomorphism: isomorphic approaches will preserve a natural one-to-one mapping between input actions and their resulting effect, whereas non-isomorphic techniques afford non-realistic interactions and can even be based on ``magical" or ``virtual" tools. Quite often, these techniques either rely on a \emph{touching} metaphor a \emph{pointing} metaphor.

As for touch-based techniques, the user must reach the target object's position to interact with it. An isomorphic example of such a technique would be to track a physical controller with six degrees of freedom and map its position and orientation to a virtual cursor, so that interaction with an object can be achieved by pressing a button when the virtual cursor collides with the target object. When the tracking zone is limited in space compared to the ``interactable" area, non-isomorphic mappings are typically used to mitigate these limitations. Examples include the Go-Go~\cite{poupyrev1996go} and PRISM~\cite{frees2005precise} techniques, that both rely on non-linear mappings between the cursor and the tracked objects' motions. 

The pointing metaphor allows to mitigate space-related limitations, since aiming at an object is sufficient to start interacting with it. Similarly, the user can reposition that object by pointing towards a target position. Typically, a tracked controller's position and orientation defines a ``laser beam" that is used to select and manipulate objects. 
Techniques based on that metaphor differ in how the position and orientation of the controller affect the laser beam, and which object(s) are selected based on it. The simplest version is often called ray-casting, where the controller defines a line segment and its intersection with the environment defines the target object or position. More complex techniques allow users to bend the line (e.g. using Bézier curves \cite{feiner2003flexible}) to mitigate limitations related to occlusion, or make use of selection volumes instead of simple rays, which can result in easier selection and can potentially allow multiple objects to be selected. The selection volume size can be static (e.g. the spotlight technique \cite{liang1994jdcad} that relies on a selection cone) or dynamic (e.g. the aperture technique \cite{forsberg1996aperture} that expands on spotlight by allowing users to control the spread angle of the cone).


\subsubsection{Speech Recognition}
In addition to the aforementioned interaction techniques, specific actions can be greatly simplified by relying on speech recognition. Since specifying an arbitrary position or selecting an existing object is easily done with a pointing metaphor, speech recognisers are often used in a multimodal context when applied to 3D selections or manipulations (e.g. the ``Put-That-There" metaphor \cite{bolt1980put}).

A plethora of speech recognition tools and techniques are available for use. Some can be used offline whereas others are based on an online service; they can either listen to the user in a continuous manner or await specific actions (e.g. a button press or an API call).
An important distinction between speech engines is whether (and how much) they restrict potential input. Free speech recognisers can output any text whereas directed dialogue \cite{pieraccini2005we} systems are limited to a set of predefined words or commands. Directed approaches can mostly be found in two forms: keyword-spotting solution that extract specific words; and grammar-based tools that produce phrases defined by specific rules.

Our use case would benefit from vocal commands such as ``Add component X" or ``Add slider with value 7", to create new (potentially valued) nodes in the graph.
Even though free speech and keyword-based approaches could be used for that purpose, they would not guarantee that a valid output is returned by the speech recogniser and would require manual parsing of that output (which sequences of words or keywords are valid, and what action they correspond to). Grammar-based engines therefore seem to be the best option as only valid vocal commands, with regards to the grammar, can be recognised. The challenge therefore moves from post-processing the result to correctly defining the rules of the grammar. The remaining of this section will consequently present the Speech Recognition Grammar Specification (SRGS). This is a W3C standard\footnote{\url{https://www.w3.org/TR/speech-grammar/#S1}} that describes a grammar format. Similarly to the grammars from compiler theory, a SRGS grammar describes a set of rules composed of tokens, using either an XML or a BNF-based (Backus-Naur Format) syntax. 

SRGS grammars can be augmented with Semantic Interpretation for Speech Recognition (SISR\footnote{\url{https://www.w3.org/TR/semantic-interpretation/}}) tags that contain ECMAScript (JavaScript) code to be executed when the corresponding grammar rule is matched. Those tags are typically used to assign values to a matched rule (e.g. a boolean value can be set to \verb|true| when the matched text is ``yes", ``ok" or ``yeah"), especially when handling numbers (e.g. saying ``three" assigns the value \verb|3| to the variable \verb|outValue|; and saying ``thousands" multiplies \verb|outValue| by 1000).

\section{Targeted actions and interaction possibilities}
\label{sec:actions-interactions}


\begin{table*}[ht]
\renewcommand{\arraystretch}{1.4}
\caption{Targeted actions and options for interaction techniques.}
\label{tab:actions-techniques}
\centering
\begin{tabular}{|c|c|c|c|c|c|c|c|c|}
	\hline
	\multicolumn{2}{|c|}{\multirow{2}{*}{\diagbox{Actions}{Techniques}}} & \multicolumn{3}{c|}{Modality}                                                                    & \multicolumn{4}{c|}{Interaction type}                                                                                       \\ \cline{3-9} 
	\multicolumn{2}{|c|}{}                                    & \multicolumn{1}{c|}{6-DoF controller} & \multicolumn{1}{c|}{Hands} & \multicolumn{1}{c|}{Speech} & \multicolumn{1}{c|}{Direct} & \multicolumn{1}{c|}{Indirect} & \multicolumn{1}{c|}{Isomorphic} & \multicolumn{1}{c|}{Non-isomorphic} \\ \hline
	\multirow{3}{*}{Node}               & Add                 & $P_1$                                    & \textcolor{gray}{$P_2$}                         & $P_1$                          & $P_1$, \textcolor{gray}{$P_2$}                      & $P_1$, \textcolor{gray}{$P_2$}                      & $P_1$, \textcolor{gray}{$P_2$}                          & \textcolor{gray}{$P_2$}                            \\ \cline{2-9} 
	& Remove              & $P_1$                                    & \textcolor{gray}{$P_2$}                         &                             & $P_1$, \textcolor{gray}{$P_2$}                      &                             & $P_1$, \textcolor{gray}{$P_2$}                          & \textcolor{gray}{$P_2$}                            \\ \cline{2-9} 
	& Move                & $P_1$                                    & \textcolor{gray}{$P_2$}                         &                             & $P_1$, \textcolor{gray}{$P_2$}                      &                             & $P_1$, \textcolor{gray}{$P_2$}                          & \textcolor{gray}{$P_2$}                            \\ \hline
	\multirow{2}{*}{Edge}               & Add                 & $P_1$                                    & \textcolor{gray}{$P_2$}                         &                             & $P_1$, \textcolor{gray}{$P_2$}                      &                             & $P_1$, \textcolor{gray}{$P_2$}                          & \textcolor{gray}{$P_2$}                            \\ \cline{2-9} 
	& Remove              & $P_1$                                    & \textcolor{gray}{$P_2$}                         &                             & $P_1$, \textcolor{gray}{$P_2$}                      &                             & $P_1$, \textcolor{gray}{$P_2$}                          & \textcolor{gray}{$P_2$}                            \\ \hline
\end{tabular}
\end{table*}

The two basic elements of a parametric design graph are nodes and edges, with each node containing any number of input and/or output ports.
In order to interact with the graph and in addition to the ability to modify parameter values, a user should be able to add, move and remove nodes. User should also be capable of adding and removing edges as well as moving and scaling the viewpoint. By viewpoint, we mean the part of the graph that is currently visible to the designer. 

Considering the devices we have at our disposal for this workshop (HTC Vive and its controllers\footnote{\url{https://www.vive.com/us/product/vive-virtual-reality-system/}}, Leap Motion\footnote{\url{https://www.leapmotion.com/}} and Kinect\footnote{\url{https://developer.microsoft.com/en-us/windows/kinect}}), Figure \ref{tab:actions-techniques} helps in visualising the exploration space i.e. which interaction techniques are, were or can be used to realise those actions. That table implies that we classify these interaction techniques based on the modality they rely on and whether they interact with the target object directly.

\subsection{First prototype: Vive controllers}
\begin{figure}[!t]
	\centering
	\includegraphics[width=3.4in]{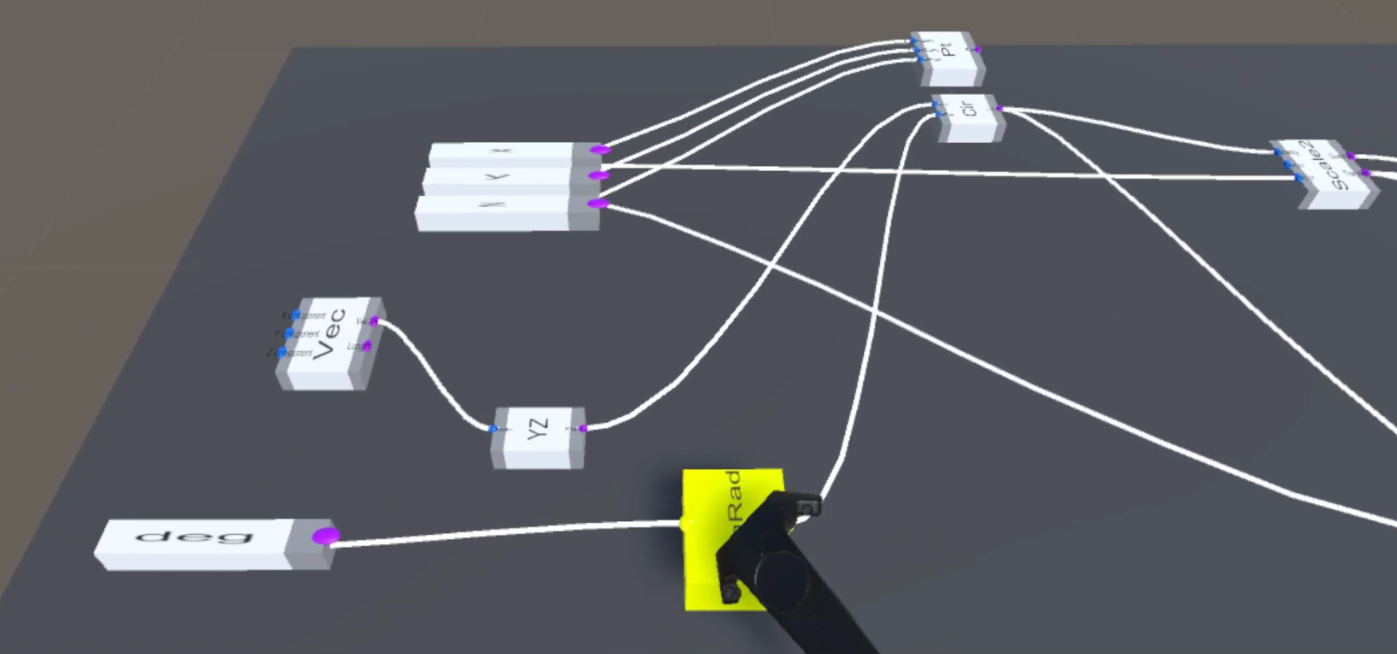}
	\caption{A user grasping a node with our first prototype.}
	\label{fig:p1}
\end{figure}

Our first prototype ($P_1$ on Table \ref{tab:actions-techniques}) relies on the default controllers provided with the HTC Vive. They are tracked with 6 degrees of freedom (DoF), meaning that their 3D position and rotation are both tracked simultaneously. For this prototype, we chose to explore an isomorphic interaction technique based on the grasping metaphor: the user simply touches the element he wants to have an interaction on, and presses a button to trigger the corresponding action. Figure \ref{fig:p1} shows a user that is about to grasp a node in $P_1$.

If that element is an edge, we simply remove it from the graph. If it is a node, we attach it to the controller (that type of interaction is often referred to as the grasping metaphor). The user can then either release it somewhere else on the graph (realising the ``move node" action) or throw it away (``remove node" action). In order to add an edge to the graph, the user simply has to select two ports. After selecting the first port and prior to selecting the second one, a temporary line between the selected port and the controller is rendered so as to give feedback to the user on the port that has been interacted with. Note that adding an edge is prevented if that edge would create a cycle in the graph (since Grasshopper can only work with acyclic graphs).

A video demonstrating this prototype is available online\footnote{\url{http://informatique.umons.ac.be/staff/Coppens.Adrien/?video=eNTERFACE2019}} and our codebase is hosted as open-source software on a GitHub repository\footnote{\url{https://github.com/qdrien/eNTERFACE-graph-architecture}}.

\subsection{Second prototype: Leap Motion}
The goal of the second prototype ($P_2$ on Table \ref{tab:actions-techniques}) is to mimic the isomorphic techniques from $P_1$ and explore non-isomorphic and indirect interaction with the Leap motion finger-tracking device. The sensor is composed of different cameras that give it the ability to scan the space above its surface, up to 60 cm away from the device.

In order to manipulate and control the basic elements of the graph-based model, we rely on gestures. Captured gestures determine what type of action to apply to these elements. We first started by recognising and categorising gestures captured from the Leap Motion, in real time. Events we recognise include ``hover", ``contact" and ``grasp" but we also can rely on a non-isomorphic gesture: ``pointing" (using raycasting).
Each of these events/gestures can be detected separately for both left and right hands, simultaneously. 

Unfortunately, $P_2$'s codebase and assets were not compatible with the framework developed for the workshop, and we consequently could not integrate the explored techniques into a fully working prototype.

\subsection{Speech recognition}
Since our prototypes are made with the Unity game engine, we can benefit from its built-in support for the Windows Speech Recognition API, that includes an XML-based SRGS grammar recogniser. We therefore defined a grammar that handles node creation. A user can indeed add a new node in the graph by saying ``Add Component \verb|type|", where \verb|type| is the node type. The list of valid node types for the grammar is dynamically generated at runtime, based on the component templates previously learned by the system (e.g. upon encountering a new component type when loading a model from a file). 

Since we sometimes need to assign a value to a newly added component (e.g. ``Add slider with value 7"), we also need to recognise numbers. Alphanumeric input in grammars is nontrivial \cite{wang2004creating} but we relied on an existing set of rules provided in the Microsoft Speech Platform SDK\footnote{\url{https://www.microsoft.com/en-us/download/details.aspx?id=27226}}.

Lastly, as users may want to assign an arbitrary text value to a node (e.g. a panel component), we had to make use of a free speech recogniser, that starts listening to user input only when specific grammar rules have been processed. In the meantime, the grammar recognition engine is paused, and it only resumes when the user stops providing free speech input.

\section{Evaluation}
\label{sec:evaluation}
\subsection{Setup}
To evaluate our prototypes, we will develop different scenarios that focus on fundamental features of the system (e.g. add a certain type of node or remove a specific edge). After an introduction to VR to limit biases related to (un)familiarity with the technology, we will ask participants to go through equivalent scenarios with our VR prototypes and question them during the session to obtain live feedback (i.e. interview/demo approach \cite{bowman2002survey}).

We will also ask participants to evaluate the usability of the system after each session, with a standard post-hoc questionnaire \cite{bowman2002survey} SUS (System Usability Scale \cite{brooke1996sus}). Combining these complimentary methods will allow us to benefit both from a deep level of live feedback and an overall evaluation of the system.

\subsection{Evaluation Criterion}
Different indicators of usability will be evaluated:
\paragraph{Completion time for a specific task}
A different input method may perform better on a particular subset of actions, and we therefore need to make sure that proper interaction techniques are available.
\paragraph{Error rate}
Quick completion only makes sense when the result matches the intended effect, so we will keep track of errors made during the experiments (e.g. the wrong edge has been deleted).
\paragraph{Intuitiveness}
Thanks to the aforementioned SUS questionnaire, we will be able to measure how easy it is for users to guess what actions will produce the intended result. Should difficulties be identified, we will take corrective actions so as to reduce the ``Gulf of Execution" \cite{LaViola3duserinterfaces2017}.

\section{Future Work}
\label{sec:future}

Since $P_2$ could not be integrated, we will have to pursue our efforts with regards to other interaction techniques before evaluating the prototypes as described in section \ref{sec:evaluation}, with architects and architectural students.
We will also adapt our work to other use cases in Software Engineering (e.g. UML diagram editing) in the near future, so as to validate the genericity of our approach.

\section{Conclusion}
After an introduction to the chosen use case's context, this report described our work on parametric modelling in VR, with a focus on the Human-Computer Interaction aspects and a multimodal approach. Even though we focused on a specific use case (parametric architectural modelling), we believe our experiments and findings are beneficial for all graph-based modelling activities in VR.
More contributions to the domain are needed to fully take advantage of VR as a visualisation technology.

\appendices




\ifCLASSOPTIONcaptionsoff
  \newpage
\fi



\bibliographystyle{IEEEtran}
\bibliography{sources}

%



%

\newpage

\begin{IEEEbiography}[{\includegraphics[width=1in,height=1.25in,clip,keepaspectratio]{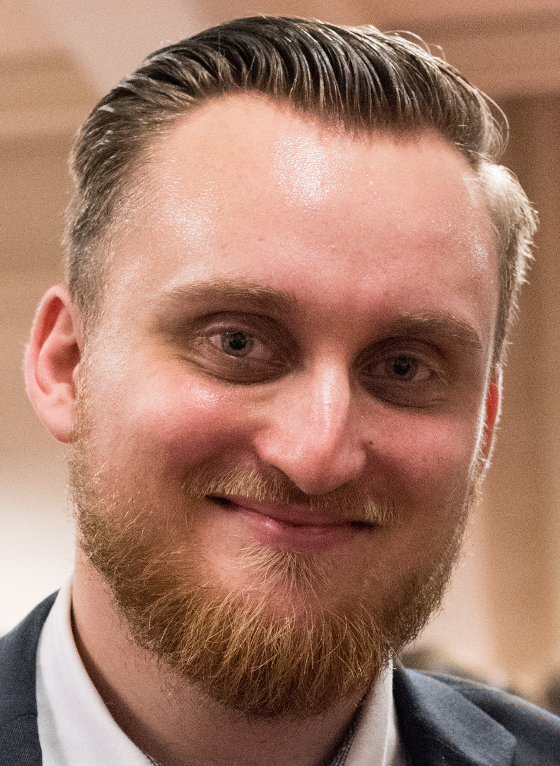}}]{Adrien Coppens}
	is a PhD student in the Software Engineering Lab at the University of Mons, in Belgium, where he also received a MSc degree in Computer Science, in 2017. His thesis is co-supervised by the Faculty of Architecture and Urban Planning since its topic is about bringing Augmented (AR) and Virtual Reality (VR) technologies in use in the context of Computer-Aided Architectural Design (CAAD), with a particular focus on parametric modelling. As relying on those technologies requires new kinds of interfaces and interaction paradigms, he is heavily interested in Human-Computer Interaction, especially within 3D immersive environments.
\end{IEEEbiography}

\begin{IEEEbiography}[{\includegraphics[width=1in,height=1.25in,clip,keepaspectratio]{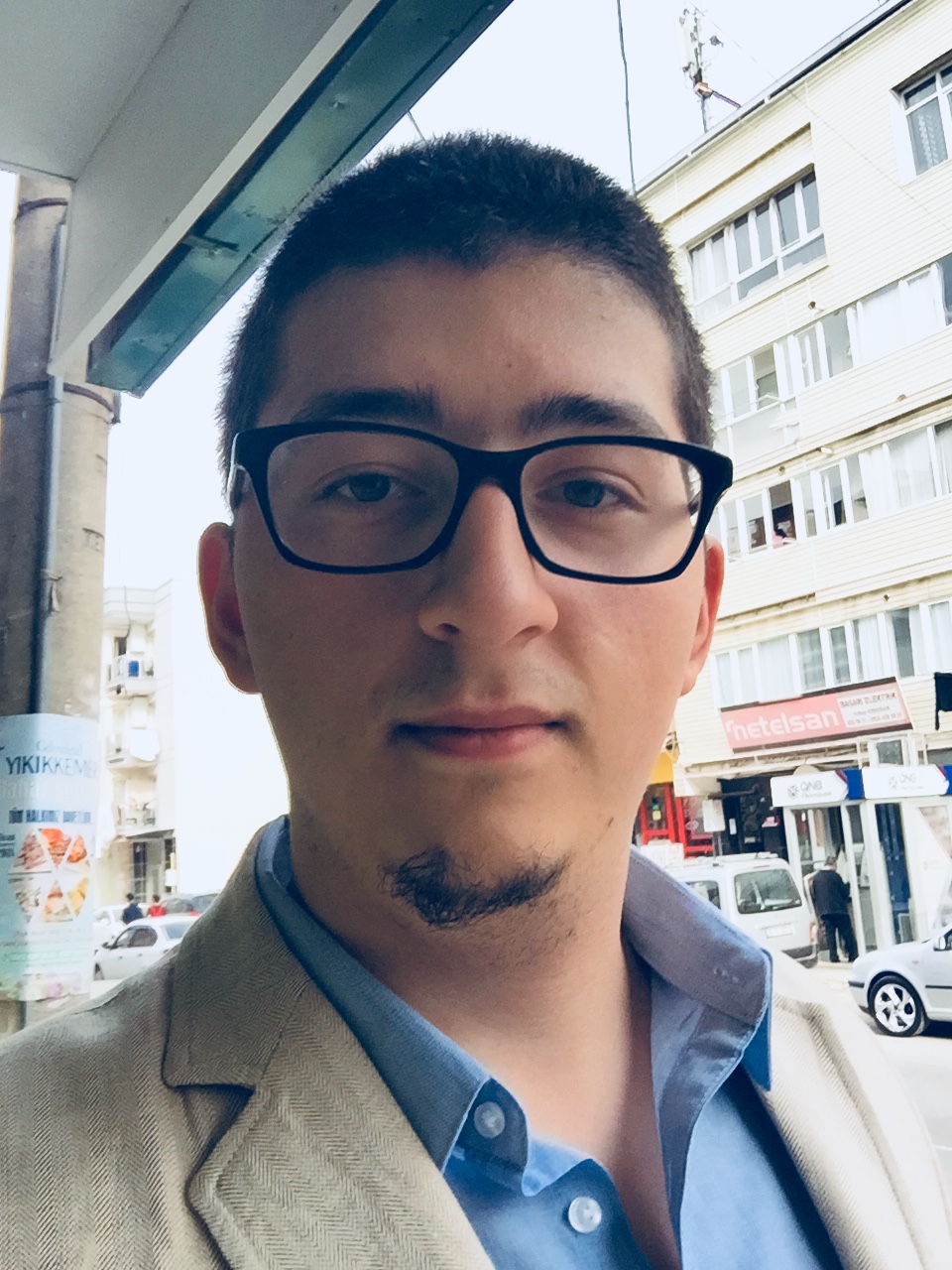}}]{Berat Bicer}
	Berat Bicer was born in Tekirdag, Turkey in 1997. He received his B. Sc. degree from Department of Computer Engineering of Bilkent University, Ankara in June 2019 and is currently pursuing his Master's Degree in Computer Engineering at Bilkent University, Ankara under supervision of Hamdi Dibeklioglu. He's currently working on multimodal deceit and online spam detection. His research interests include computer vision, human behavior analysis, and pattern recognition.
\end{IEEEbiography}

\begin{IEEEbiography}[{\includegraphics[width=1in,height=1.25in,clip,keepaspectratio]{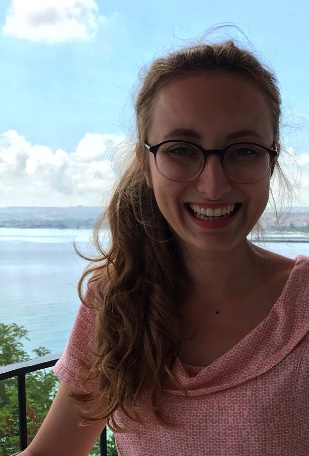}}]{Naz Yilmaz}
	Naz Buse Yilmaz received her BSc. Degree in Interior Architecture from Istanbul Technical University (ITU), in 2018. She started her graduate studies at Cognitive Science Program, Bogazici University and currently MSc. Cognitive Science student at Graduate School of Informatics, Middle East Technical University. Her research interests include perception, visual and spatial cognition and human computer interaction.
\end{IEEEbiography}

\begin{IEEEbiography}[{\includegraphics[width=1in,height=1.25in,clip,keepaspectratio]{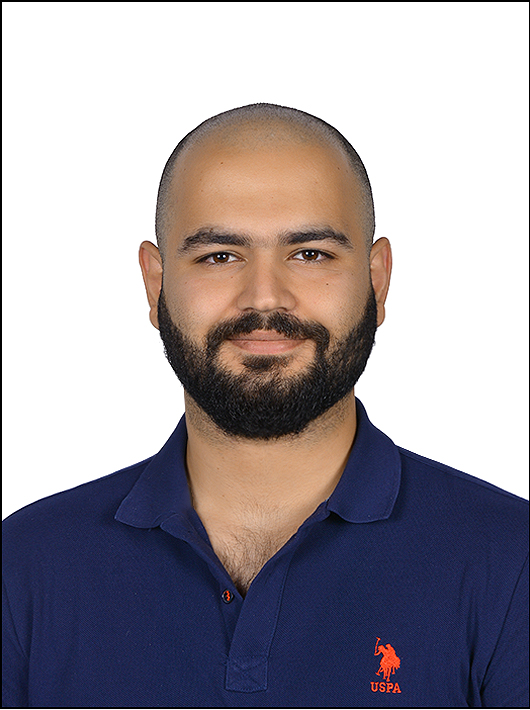}}]{Serhat Aras}
	Serhat Aras was born in Ankara, Turkey in 1996. He received his Bachelor’s Degree in Computer Science from Bilkent University, Ankara, in 2019. Currently, he is pursuing a career based on Software Development, Machine Learning and Computer Vision. He is also a researcher at Bilkent University under the supervision of Hamdi Dibeklioglu. His research interests include Pattern Recognition, Computer Vision, Human Behavior Analysis, Gesture Recognition and Edge Computing.
\end{IEEEbiography}
\vfill



%




\end{document}